\documentclass{article}
\usepackage{graphicx} %
\usepackage{subcaption}
\usepackage{amsmath}
\usepackage{hyperref}
\usepackage{algorithm, algorithmic}

\usepackage[numbers,sort&compress]{natbib}
\usepackage{amssymb}

\usepackage{authblk}
\title{Ensemble Principal Component Analysis}
\author[1]{Olga Dorabiala}
\author[1]{Aleksandr Aravkin}
\author[1,2]{J. Nathan Kutz}
\affil[1]{Department of Applied Mathematics, University of Washington, Seattle, WA, USA}
\affil[2]{Department of Electrical and Computer Engineering, University of Washington, Seattle, WA, USA}
\date{}                     
\begin{document}

\maketitle
\begin{abstract}
Efficient representations of data are essential for processing, exploration, and human understanding, and {\em Principal Component Analysis} (PCA) is one of the most common dimensionality reduction techniques used for the analysis of large, multivariate datasets today. Two well-known limitations of the method include sensitivity to outliers and noise and no clear methodology for the uncertainty quantification of the principle components or their associated explained variances. Whereas previous work has focused on each of these problems individually, we propose a scalable method called {\em Ensemble PCA} (EPCA) that addresses them simultaneously for data which has an inherently low-rank structure. EPCA combines boostrapped PCA with $k$-means cluster analysis to handle challenges associated with sign-ambiguity and the re-ordering of components in the PCA subsamples. EPCA provides a noise-resistant extension of PCA that lends itself naturally to uncertainty quantification. We test EPCA on data corrupted with white noise, sparse noise, and outliers against both classical PCA and Robust PCA (RPCA) and show that EPCA performs competitively across different noise scenarios, with a clear advantage on datasets containing outliers and orders of magnitude reduction in computational cost compared to RPCA. 
\end{abstract}

\section{Introduction}
Across the engineering, physical, biological and social sciences, data science methods are a dominant paradigm for the processing, exploration, and understanding of emerging big data. Indeed, modern sensor technologies are revolutionizing the automated collection of large data sets, which must be processed in order to aid in human understanding and decision making.  Critical for such interpretability is the generation of low-dimensional feature spaces, or dominant patterns, which can be extracted from high-dimensional and multivariate data streams~\cite{james2013introduction,wright2022high}.  One of the most successful methods for extracting dominant features in data is {\em principle component analysis} (PCA), which provides a characterization of dominant correlated activity using the underling mathematical algorithm of the {\em singular value decomposition} (SVD). The SVD has been so prolific, that it has been independently developed for dimensionality-reduction and feature extraction across a number of fields, each with their own name, including {\em proper orthogonal decomposition} (POD)~\cite{berkooz1993proper,benner2015survey,kutz2013data}, {\em empirical orthogonal functions} (EOFs)~\cite{lorenz1956empirical,hannachi2007empirical}, Karhunen-Lo\'eve expansion~\cite{dony2001karhunen,kirby1990application}, and the Hoteling transform~\cite{hotelling1938transformation}. Computational algorithms of the early 1970s~\cite{golub1971singular} allowed for the robust and mature extraction of PCA modes from data, which has made PCA a classical data analysis technique. However, PCA is sensitive to noise and outliers, and the extraction of PCA modes does not come with any uncertainty quantification (UQ) metrics. We address both of these issues in our innovation of {\em Ensemble PCA}, which builds upon classic PCA in order to stabilize PCA models and provide clear UQ metrics for a given data set.

The motivations for dimensionality reduction are extensive. Often, the features of high-dimensional data exhibit partial redundancy and dependency. 
Beyond reducing redundancy, extracting the ``most important" features of a data matrix that best summarize the information contained in a signal and removing irrelevant features helps practitioners both to work with limited computational resources and to understand the underlying data structure. Further, projecting high-dimensional data into 2D or 3D space can be extremely beneficial for human visualization. Too many features can complicate data analysis and visualization, and dimensionality reduction helps avoid these pitfalls. There are two main classes of dimensionality reduction techniques: linear and nonlinear. At their core, linear techniques use linear transformations to shift and stretch data. Examples include SVD, PCA, and Fisher’s {\em linear discriminant analysis} (LDA). These methods are a cornerstone of analyzing data due to their simple geometric interpretations and typically attractive computational properties ~\cite{cunningham2015linear}. They are particularly useful when the data lies in a linear subspace and where the original variables are replaced by a smaller set of underlying variables. Nonlinear techniques involve more complicated data transformations and include {\em t-distributed stochastic neighbor embedding} (t-SNE)~\cite{van2008visualizing}, Isomap~\cite{balasubramanian2002isomap}, and autoencoders~\cite{vincent2008extracting,goodfellow2016deep}. Nonlinear methods are generally more powerful than their linear counterparts, but can be slow to optimize and many are non-deterministic, meaning they get different, locally-optimal solutions on each run~\cite{silva2002global}.

The focus of this work is linear dimensionality reduction, particularly PCA~\cite{jolliffe2005principal}, as it is arguably the most common dimensionality reduction technique used for the analysis of large, multivariate datasets today. Innovations in randomized methods now allow PCA to work at massive scales~\cite{halko2011finding,drineas2016randnla,erichson2016randomized}. 
Extensions of PCA include Kernel PCA, which runs the algorithm on a feature space determined by the kernel~\cite{mika1998kernel}, probabilistic PCA ~\cite{tipping1999probabilistic}, which 
is targeted for data with missing entries, and sparse PCA ~\cite{zou2006sparse}, which seeks to summarize data using combinations of only a few input variables.

Regardless of extension, PCA has two well-known shortcomings. The first is sensitivity to outliers, where entire rows of the data matrix may be contaminated, and sparse noise, where individual entries of the data matrix may be affected. The second drawback is that there is no clear methodology for UQ.
Previous work has focused on each of these problems individually. Robust PCA extends PCA to the sparse noise domain~\cite{bouwmans2014robust}, and bootstrapping has been explored as a method for estimating sampling variability~\cite{fisher2016fast}. To our knowledge, there has not been an extensive analysis of bootstrapping for the purpose of developing a noise-resistant PCA. Gabrys et. al~\cite{gabrys2006outlier} suggested that statistical resampling could be applied to PCA to recover the true components of datasets corrupted with large outliers, but failed to test this concept on a sufficient number of datasets and wholly ignored UQ. 
Drawing inspiration from~\cite{gabrys2006outlier} we propose Ensemble Principal Component Analysis (EPCA), which ensembles bootstrapping with $k$-means cluster analysis to create a noise-resistant approach. We test EPCA against RPCA and standard PCA on seven datasets corrupted with sparse noise, white noise, and outliers. We show that EPCA achieves maximum performance on datasets with outliers and performs competitively on datasets with other types of noise. EPCA also naturally supports uncertainty quantification and provides an orders of magnitude reduction in computational cost compared to RPCA.


\section{Related Work}

PCA operates as follows. Let ${{\bf X} = [{\bf x}_1, {\bf x}_2, \cdots, {\bf x}_m] \in \mathbb{R}^{N\times m}}$ be a data matrix with $N$ samples and $m$ features. Define the mean-centered matrix as ${\overline{{\bf X}} = {\bf X} - \mu_{\bf X}}$, where each feature is centered by its mean. The sample covariance matrix ${{\bf C} \in \mathbb{R}^{m \times m}}$ of $\overline{{\bf X}}$ is then defined as 
\begin{equation}
    {\bf C} =\frac{\overline{{\bf X}}^T\overline{{\bf X}}}{n-1}.
\end{equation} 
%
Since ${\bf C}$ is symmetric, it can be diagonalized as 
\begin{equation}
    {\bf C} = {\bf V } {\bf \Lambda}{\bf V}^*,
\end{equation}
\noindent where ${\bf V}$ contains the eigenvectors of ${\bf C}$ and ${\bf \Lambda}$ is a diagonal matrix containing the eigenvalues $\lambda_i$ on the diagonal, ordered by decreasing magnitude. The principal components of ${\bf X}$ are defined as $\overline{{\bf X}}{\bf V}$. The principal component associated with the largest eigenvalue projects the data onto the direction of greatest variance, while the eigenvalue measures the amount of information that can be explained by that projection ~\cite{van2004pca}. The $d$ largest eigenvalues are referred to as explained variance.




The results of PCA can be unreliable under data corruption. We identify three main classes of noise: white noise, sparse noise, and outliers. Under white noise, all entries of a data matrix ${\bf X}$ are slightly perturbed. 
Under sparse noise, a small subset of the elements of ${\bf X}$ are corrupt with some probability $p$, and with outliers, multiplicative noise is applied to a given percentage of rows of ${\bf X}$.
PCA will break down if even one entry of ${\bf X}$ is grossly corrupted ~\cite{lin2010augmented}.

The sensitivity of PCA has motivated the development of noise-resistant extensions. Directions of research include the use of robust estimators of scatter as opposed to sample covariance~\cite{van2004pca} and the development of Robust Principal Component Analysis (RPCA). RPCA sets the standard for dimensionality reduction under sparse noise, and is generally applied for foreground detection~\cite{bouwmans2014robust}. Mathematically, RPCA assumes we observe a data matrix ${{\bf X} \in \mathbb{R}^{N \times m}}$ of the following form ${{\bf X} = {\bf L}_0 + {\bf S}_0 + {\bf Z}_0}$, where ${\bf L}_0$ is low-rank, ${\bf S}_0$ is sparse, and ${\bf Z}_0$ is a noise term containing i.i.d. noise on each entry ~\cite{candes2011robust, zhou2010stable}. Generally, RPCA recovers ${\bf L}_0$ and ${\bf S}_0$ by solving the optimization problem in \eqref{eq:rpca} 

\begin{equation}
    \min_{{\bf L},{\bf S}} ||{\bf L}||_* + \alpha ||{\bf S}||_1,
    \label{eq:rpca}
\end{equation}
subject to some constraint on the ${\bf L}$ and ${\bf S}$ matrices, such as ${{\bf X} = {\bf L} + {\bf S}}$ or ${||{\bf X} - {\bf L} - {\bf S}|| \leq \delta }$, where $\delta$ is some tuning parameter and the norm can be customized~\cite{bouwmans2014robust}. In Expression \eqref{eq:rpca}, $||\cdot||_*$ denotes the nuclear norm, $||\cdot||_1$ denotes the sum of the absolute values of matrix entries, and $\alpha>0$ is a tuning parameter controlling the regularization of ${\bf S}$. 

Equation \eqref{eq:rpca} is referred to as the Principal Component Pursuit (PCP) problem, and a handful of methods have been suggested for solving PCP ~\cite{bouwmans2014robust, lin2010augmented}. PCP can recover ${\bf L}$ and ${\bf S}$ exactly ~\cite{candes2011robust}, but is limited to the case where the low-rank component is exactly low-rank and the sparse component is exactly sparse. In addition, existing algorithms for solving PCP are computationally expensive and require extensive parameter tuning ~\cite{bouwmans2014robust}. Another set of extensions to PCA focus on missing data, where entries of ${\bf X}$ are deleted at random. Most approaches specialized for dealing with missingness focus on data imputation ~\cite{josse2012handling, zhu2019high}. However, missingness can be considered a sub-category of sparse noise, and methods such as RPCA continue to work in this context. 

Another drawback of PCA is that it offers no clear method for estimating the sampling variablity of its descriptive outputs. Though some work has explored analytical, asymptotic confidence intervals (CIs) for principal components, the methods are often either computationally infeasible or require strong assumptions on the data~\cite{fisher2016fast}. An alternative approach is using bootstrap-based CIs. Bootstrapping is a way to assess the variability of a statistic of interest through random sampling with replacement. It does not assume any distribution for the estimates of the uncertainties, and can be applied to most statistics~\cite{babamoradi2013bootstrap}. 

In the context of PCA, boostrapping 
estimates the variability of PCA across different samples of the population~\cite{fisher2016fast}. From a theoretical standpoint, boostrapping is generally not useful unless we are interested in inference concerning only a few large eigenvalues, which are well-separated from the bulk and of multiplicity one~\cite{karoui2016bootstrap}, i,e. data of inherently low-rank. Though bootstrapping has been applied for the uncertainty quantification of PCA, very little work has been done using bootstrapping as a method for robustification. 

Gabrys et. al~\cite{gabrys2006outlier} suggested using statistical resampling as a way to recover the principal components of a data matrix corrupted with outliers, and taking inspiration from his work, we propose EPCA, which ensembles bootstrapping PCA with $k$-means clustering to create a method for the dimensionality reduction and analysis of low-rank, noisy data that also lends itself to uncertainty quantification.  By using $k$-means to aggregate the output of boostrapped samples, we circumvent the challenges associated with component re-ordering and sign ambiguity. We test the performance of EPCA against classical PCA and Robust PCA, with respect to runtime and accuracy, on datasets corrupted with sparse noise, white noise, and outliers. 

\section{Ensemble Principle Component Analysis (EPCA)}
Given a data matrix ${{\bf X} \in \mathbb{R}^{N\times m}}$, EPCA samples $B$ bags of size $n$ at random with replacement. PCA is run on each of the $B$ samples, the principal components are stored in a matrix ${\bf P}^{(j)}$, and the eigenvalues are stored in a matrix ${{\bf \Lambda}^{(j)}}$ for ${j \in [1,B]}$. The goal is to summarize the results of our $B$ samples to output $d$ dominant modes. 

\begin{figure}
    \centering \includegraphics[width=\linewidth]{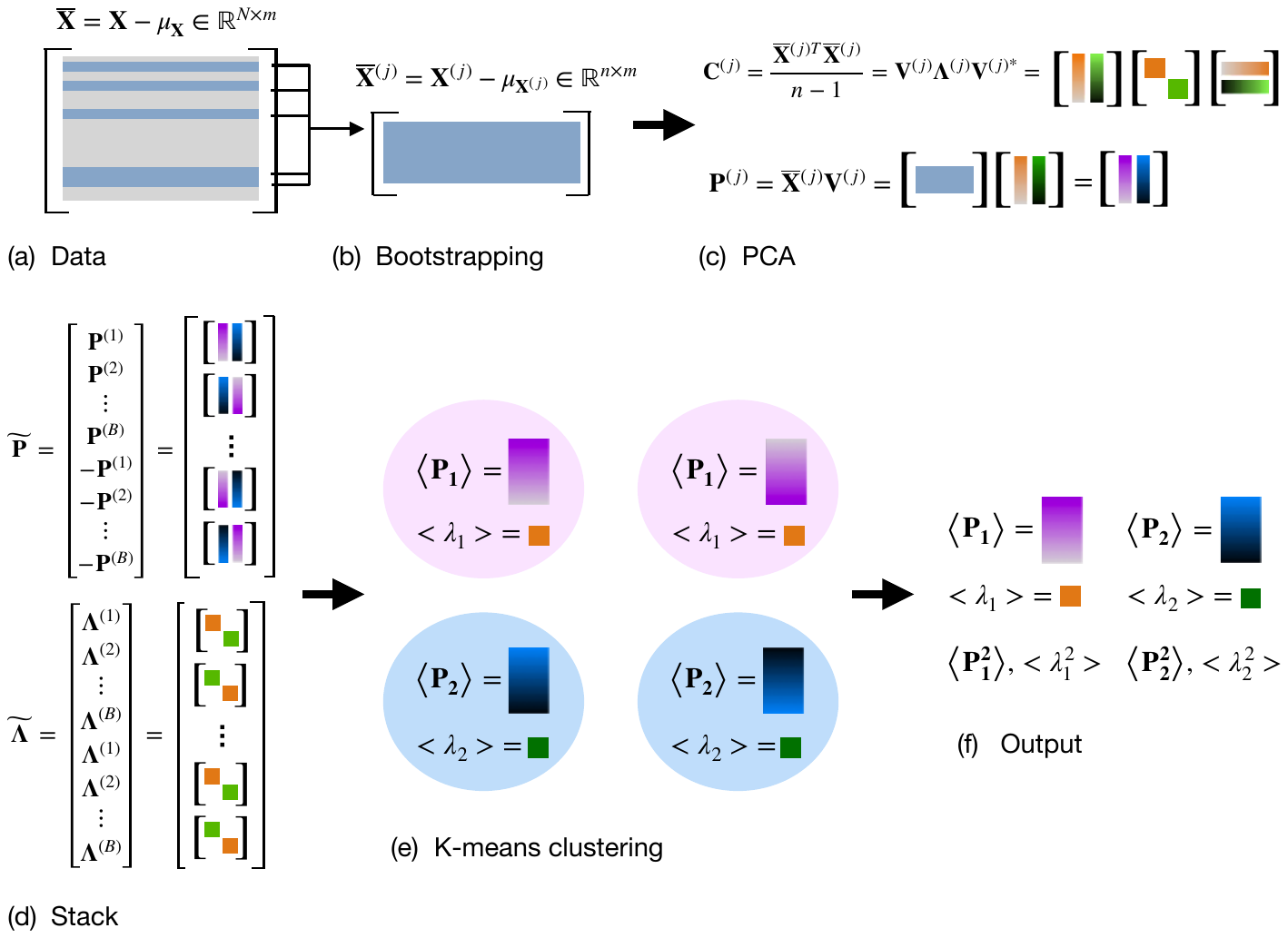}
    \caption{Ensemble Principal Component Analysis (EPCA). Given a data matrix with inherently low-rank structure, we sample $B$ bags of size $n$ at random with replacement. We run PCA and store $d$ principal components ${\bf P}^{(j)}$ and their corresponding eigenvalues ${\bf \Lambda}^{(j)}$ for each bag. We create ${\bf \widetilde{P}}$ by stacking all components, along with their reflections to account for rotational variability. We also stack all eigenvalues in ${\bf \widetilde{\Lambda}}$, in accordance with the order in ${\bf \widetilde{P}}$. Next, we run $k$-means clustering on ${\bf \widetilde{P}}$ with $2d$ clusters. }
    \label{fig:EPCA}
\end{figure}

Two challenges are that there is rotational variability in the principal components found by PCA and that the identified components can be re-ordered in the subsamples~\cite{timmerman2007estimating}. Most boostrapping PCA approaches tackle the first issue by using a Procrustean rotation to match the boostrap PCs to the PCs obtained by running PCA on the entire dataset~\cite{milan1995application} or by rotating the PCs towards some pre-specified target matrix ${\bf T}$ ~\cite{timmerman2007estimating}. Instead, we create the matrix 

\begin{equation}
    \widetilde{{\bf P}} = \begin{bmatrix} {\bf P}^{(1)} \\ {\bf P}^{(2)} \\ \vdots \\ {\bf P}^{(B)} \\ -{\bf P}^{(1)} \\ -{\bf P}^{(2)} \\ \vdots \\ -{\bf P}^{(B)} \end{bmatrix}
\end{equation}
by stacking all of the principal components found in the bags and their reflections. In this way, every principal component is stored along with its reflection, regardless of initial orientation. We also stack the corresponding eigenvalues accordingly, creating 

\begin{equation}
    \widetilde{{\bf \Lambda}} = \begin{bmatrix}{\bf \Lambda}^{(1)} \\ {\bf \Lambda}^{(2)} \\ \vdots\\ {\bf \Lambda}^{(B)} \\ {\bf \Lambda}^{(1)} \\ {\bf \Lambda}^{(2)} \\ \vdots \\ {\bf \Lambda}^{(B)}\end{bmatrix}.
\end{equation}
The next step is to run $k$-means clustering on $\widetilde{{\bf P}}$ to output $2 d$ clusters. This approach automatically clusters components that are oriented in the same direction and avoids any challenges associated with re-ordering. We use the normalized cluster centers of our $d$ rotationally unique clusters as our predicted principal components, and the averages of the eigenvalues associated with the members of each cluster as our predicted eigenvalues. We order our final predicted components according to the magnitude of the average predicted eigenvalues. EPCA is visualized in Figure \ref{fig:EPCA} and explained in  Algorithm \ref{alg:EPCAalg}.

\begin{algorithm}
\caption{Ensemble Principal Component Analysis}\label{alg:EPCAalg}
\begin{algorithmic}
\STATE (a) Mean center the data. {$\overline{\bf X} = {\bf X} - \mu_{\bf X}$} 
\STATE (b) {Select $B$ bags of size $n$ with replacement to create $\overline{X}^{(j)}$ for $j \in [1,B]$.}
\STATE (c) Run PCA on each of the $B$ bags to output $d$ eigenvalues ${\bf \Lambda}^{(j)}$ and $d$ principal components ${\bf P}^{(j)}$
\STATE (d) Stack all ${\bf P}^{(j)}$ and their reflections to create $\widetilde{\bf P}$. Stack corresponding ${\bf \Lambda}^{(j)}$ to create $\widetilde{\bf \Lambda}$. 
\STATE (e) Run $k$-means clustering on $\widetilde{\bf P}$ with $2d$ clusters to cluster principal components oriented in the same direction. 
\STATE (f) Output the directionally unique $d$ average principal components, their corresponding average eigenvalues, and the variances of both.
\end{algorithmic}
\end{algorithm}

\section{Uncertainty Quantification} 
\label{sec:UQ}

Like all other boostrapping PCA methods ~\cite{babamoradi2013bootstrap, fisher2016fast, timmerman2007estimating, milan1995application}, EPCA lends itself naturally to uncertainty quantification. There are many approaches to estimate a confidence interval (CI) from the boostrap distribution, but we focus on the percentile method~\cite{timmerman2007estimating}.

\begin{figure}
     \centering
     \begin{subfigure}[b]{\textwidth}
         \centering
         \includegraphics[width=\textwidth]{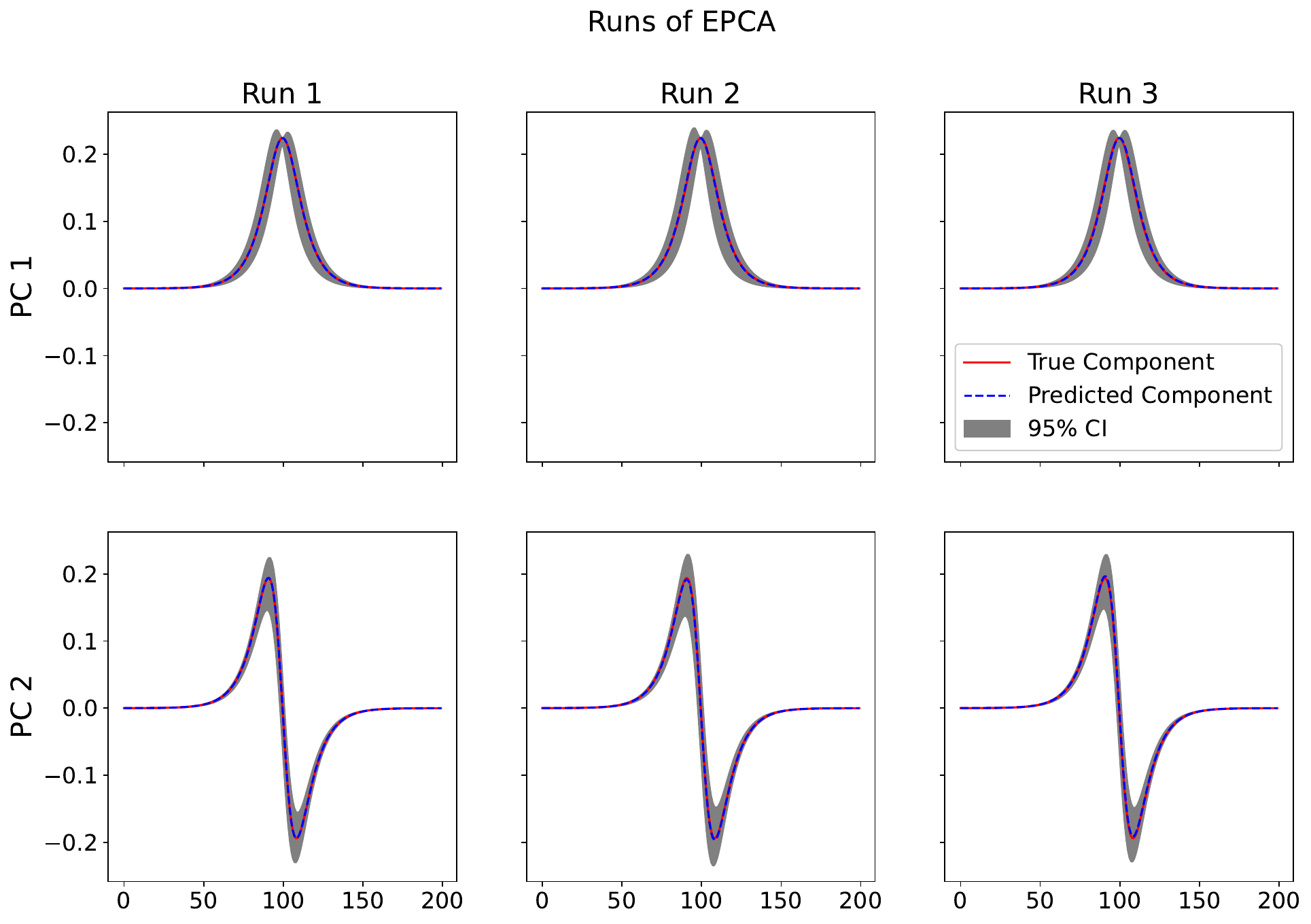}
         \caption{}
         \label{fig:UQ_original_data}
     \end{subfigure}
     \hfill
     \begin{subfigure}[b]{\textwidth}
         \centering
         \includegraphics[width=\textwidth]{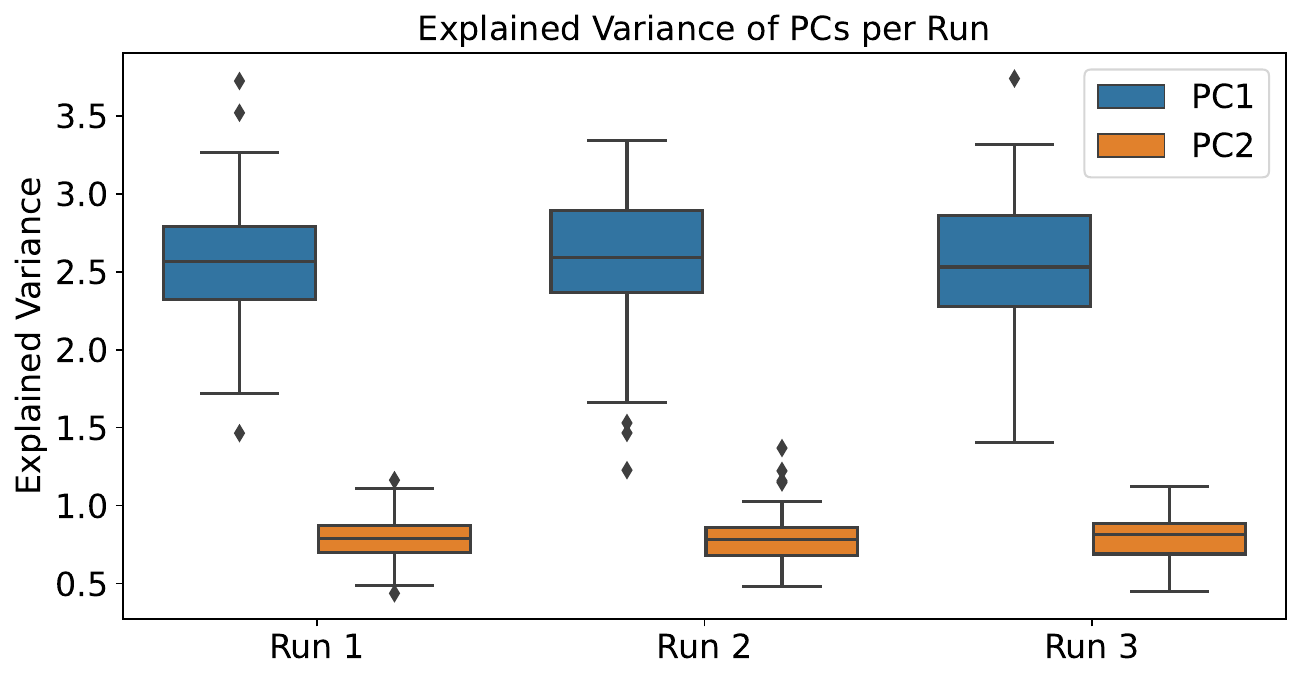}
         \caption{}
         \label{fig:UQ_original_ev}
     \end{subfigure}
    \caption{Uncertainty Quantification of three runs of EPCA on wave data. (a) The $95\%$ Confidence Intervals (CIs) associated with the top two principal components. The CIs are tight, suggesting high confidence in the predicted components. (b) Boxplots of the distribution of the explained variance of each of the principal components. The true eigenvalues of the covariance matrix are contained within the interquartile ranges (IQRs) of all three trials.}
    \label{fig:UQ_original}
\end{figure}

\begin{figure}
     \centering
     \begin{subfigure}[b]{\textwidth}
         \centering
         \includegraphics[width=\textwidth]{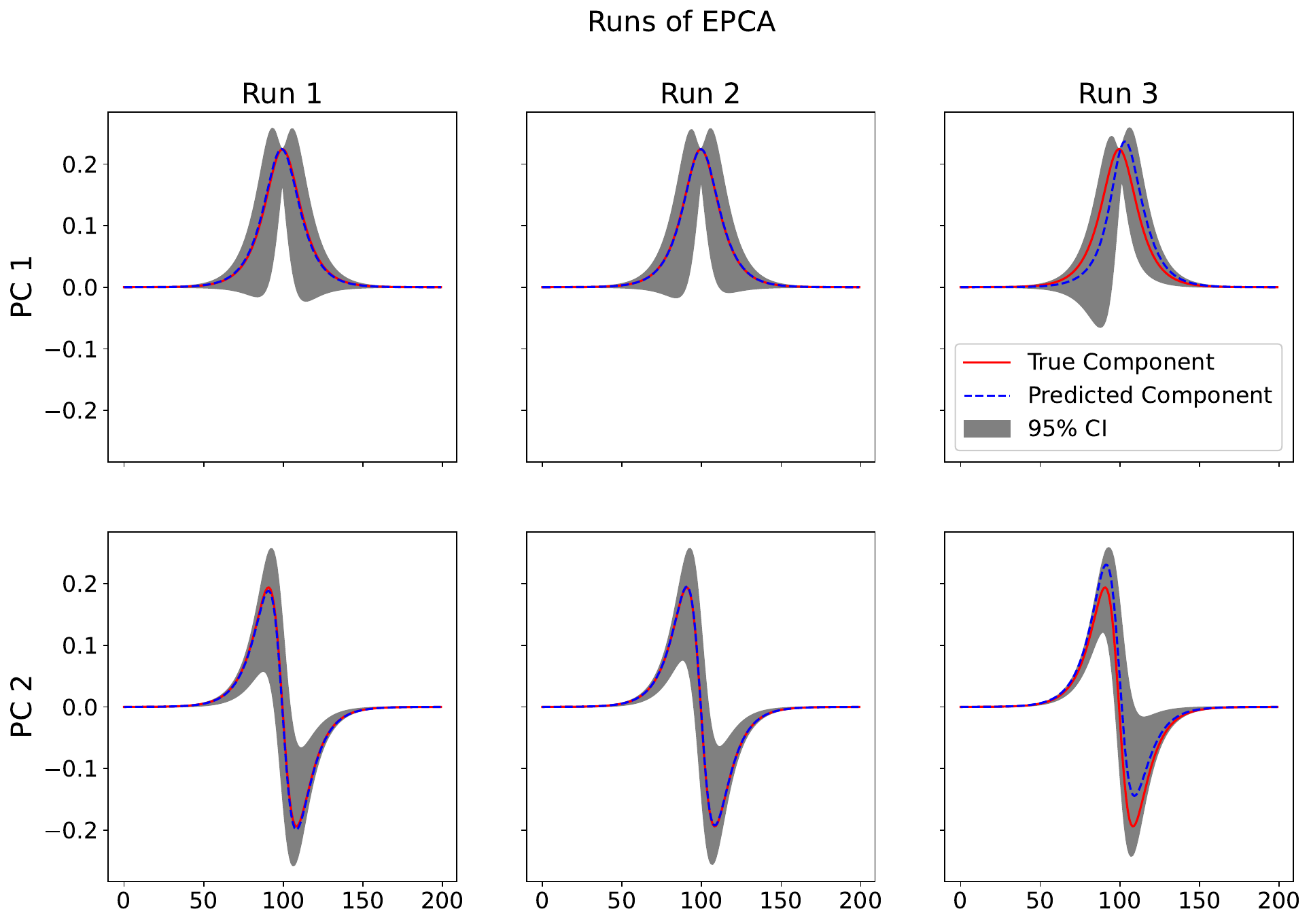}
         \caption{}
         \label{fig:UQ_outlier_data}
     \end{subfigure}
     \hfill
     \begin{subfigure}[b]{\textwidth}
         \centering
         \includegraphics[width=\textwidth]{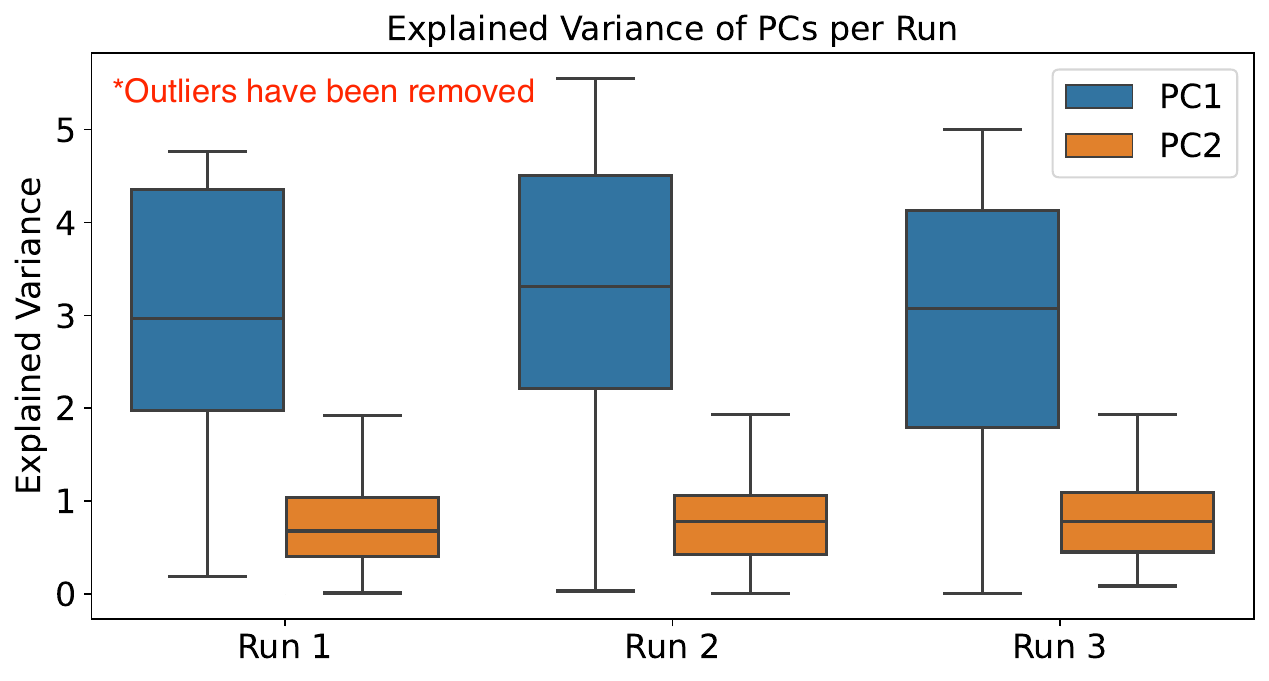}
         \caption{}
         \label{fig:UQ_outlier_ev}
     \end{subfigure}
    \caption{Uncertainty Quantification of three runs of EPCA on wave data corrupt with $5\%$ outliers of scale $10$. (a) The $95\%$ Confidence Intervals (CIs) associated with the top two principal components. The CIs are wider than that of the clean data. (b) Boxplots of the distribution of the explained variance of each of the principal components. The true eigenvalues of the covariance matrix are contained within the interquartile ranges (IQRs) of all three trials, but the IQRs are skewed upward and wider than those from the clean data. Gross outliers have been removed from the boxplots for visualization purposes.}
    \label{fig:UQ_outlier}
\end{figure}

Figures \ref{fig:UQ_original} and \ref{fig:UQ_outlier} show UQ for three runs of EPCA on a clean dataset and the same dataset containing $5 \%$ outliers of scale $10$, respectively. We show the results of three runs, since the results of EPCA vary based on the random bootstrap samples. Figures \ref{fig:UQ_original_data} and \ref{fig:UQ_outlier_data} show the $95\%$ CIs for the principal components. The tightness of our CIs correlates with the level of confidence in our output. We note that the CIs are significantly tighter for the clean data, and even though there is greater uncertainty on the corrupted data, EPCA is still able to closely identify the true components in two of the three pictured runs. Figures \ref{fig:UQ_original_ev} and \ref{fig:UQ_outlier_ev} show the distributions of the respective eigenvalues, which tell us the amount of variance explained by each of the principal components. On both the original and corrupt data, the interquartile ranges (IQR) of the variances capture the true values, but on the corrupted data, the IQRs are skewed higher and slightly wider. Gross outliers have been removed from the boxplots in Figure \ref{fig:UQ_outlier_ev}. Even though the explained variances are more difficult to capture on the outlier data, they at least tell us about the order of the principal components and their separation. In practice, the eigenvalues of the covariance matrix are not used in PCA's projection of data into a lower-dimensional space, so their explicit values are not as important as the principal components themselves.



\section{Experiments}
We test EPCA against RPCA (solved using Augmented Lagrange Multipliers~\cite{lin2010augmented}) and classical PCA on seven datasets with four types of added noise: sparse, Gaussian white, uniform white, and outliers. We include three small, three medium, and one large dataset in our analysis. Our base datasets, in order of increasing size, are iris~\cite{scikit-learn}, wine~\cite{scikit-learn}, breast cancer Wisconsin (WBC)~\cite{scikit-learn}, a synthetic wave, the digits 0 and 1 from  MNIST~\cite{deng2012mnist}, and sea surface temperature (SST)~\cite{sstdata}.

\subsection{Parameter Selection}
The first type of noise is sparse noise, where each entry in the data matrix ${\bf X}$ is corrupt with probability $p$. The value of corrupt entries is set to $c$. As the number of features grows, the number of corrupted entries will also grow. 
The next two types of noise are types of white noise, where noise from either the uniform distribution with mean $0$ and variance $v$ or the Gaussian distribution with mean $0$ and variance $v$ are added to ${\bf X}$. We ensure the variance is small with respect to the dominant singular value of ${\bf X}$, as this is a level of corruption under which classical PCA should still perform well~\cite{ding2011bayesian}. In practice, we set ${v = \frac{\sigma_1}{f}}$, where $\sigma_1$ is the dominant singular value of ${\bf X}$ and $f$ is some variance divisor. Finally, we create outliers by multiplying a randomly selected $s\%$ of the rows of ${\bf X}$ by an outlier scale of $S$. We expect RPCA to perform best on sparse data, PCA to perform well even in the presence of low-variance white noise~\cite{ding2011bayesian}, and EPCA to perform best on outlier data. 

For all datasets, we perform EPCA with $B=100$ bags, but vary the size $n$ of the bags. We take $n$ to be larger on datasets with white noise and sparse noise and smaller on datasets with outliers. Since white noise is applied to all entries of ${\bf X}$ and sparse noise will impact most entries as ${\bf X}$ becomes higher-dimensional, choosing larger bag sizes $n$ helps ``mute" the impact of the noise. Contrastingly, since outliers impact only $s\%$ of entries, choosing smaller bags helps prevent the selection of an outlier. In RPCA, the parameter controlling the extent of the regularization on the sparse part of ${\bf X}$ is set to ${\alpha = 0.20}$. The remaining parameters for the Augmented Lagrange Multipliers algorithm used to solve RPCA are set to their default values.  

\subsection{Evaluation}
Given an initial data matrix ${\bf X}$, we calculate the true PCA modes using classical PCA. We then corrupt ${\bf X}$ 
and run PCA, RPCA, and EPCA on each of the corrupted datasets and quantify the percent relative error for the components using Equation \ref{eq:rel error}. 
\begin{equation}
    \% \textrm{Relative Error} = \frac{||t - p||_2}{||t||_2} \times 100,
    \label{eq:rel error}
\end{equation}
where $t$ are the true components and $p$ the predicted components. As we saw in Section \ref{sec:UQ}, the eigenvalues of the covariance matrix in EPCA are skewed on noisy data. However, the eigenvalues are not involved in PCA's low-dimensional mapping, only in ordering the predicted components. Therefore, we consider only the error in the principal components in our experiments. 

\subsection{Various Levels of Noise}
Our first set of experiments addresses how PCA, EPCA, and RPCA respond to various levels of noise. For sparse noise, we vary both the probability $p$ of an entry being corrupt and the scale $c$ of the corruption. For both normal and uniform white noise, we vary the scale of the variance divisor $f$. Finally, for outliers, we vary both the percentage $s$ of corrupt rows and the scale $S$. 
We take each of six datasets, excluding SST, and randomly add noise of a given level five times, resulting in $30$ datasets per noise level. We do not include the SST dataset in this analysis, as RPCA is unable to produce a result within $120$s of runtime.
Since the results of PCA and RPCA are deterministic,
while the results of EPCA are stochastic due to the randomness in the bagging procedure, 
we run PCA and RPCA once and EPCA five times on each of the datasets. We average each method's respective relative errors together to get an average percent relative error for every level of corruption.

\subsection{Fixed Levels of Noise}
In the second set of experiments, we compare the performance of PCA, RPCA, and EPCA on datasets with fixed levels of corruption to test variability in performance. For sparse noise, we set the probability of an entry being corrupt to $p=0.01$ and the scale of the corruption to $c=2$, for both Gaussian and uniform white noise, we set the variance divisor to $f = 1000$, so that the variance is $v = \frac{\sigma_1}{1000}$, and for outliers, we corrupt $s = 5$ percent of rows with scale $S = 5$. For each of our seven datasets and each type of noise, we repeat random corruption and evaluation $100$ times, resulting in $700$ runs of each method per noise category. In our analysis, we consider only the output of a single run of EPCA. We return boxplots of percent relative error over the $100$ trials. Once again, RPCA is unable to produce a result on the SST dataset before a timeout at $120$s; therefore, RPCA's boxplots do not include performance metrics on SST. We also carry out a runtime comparison among the three methods.


\section{Results}
\subsection{Various Levels of Noise}
\begin{figure}
    \centering \includegraphics[width=\linewidth]{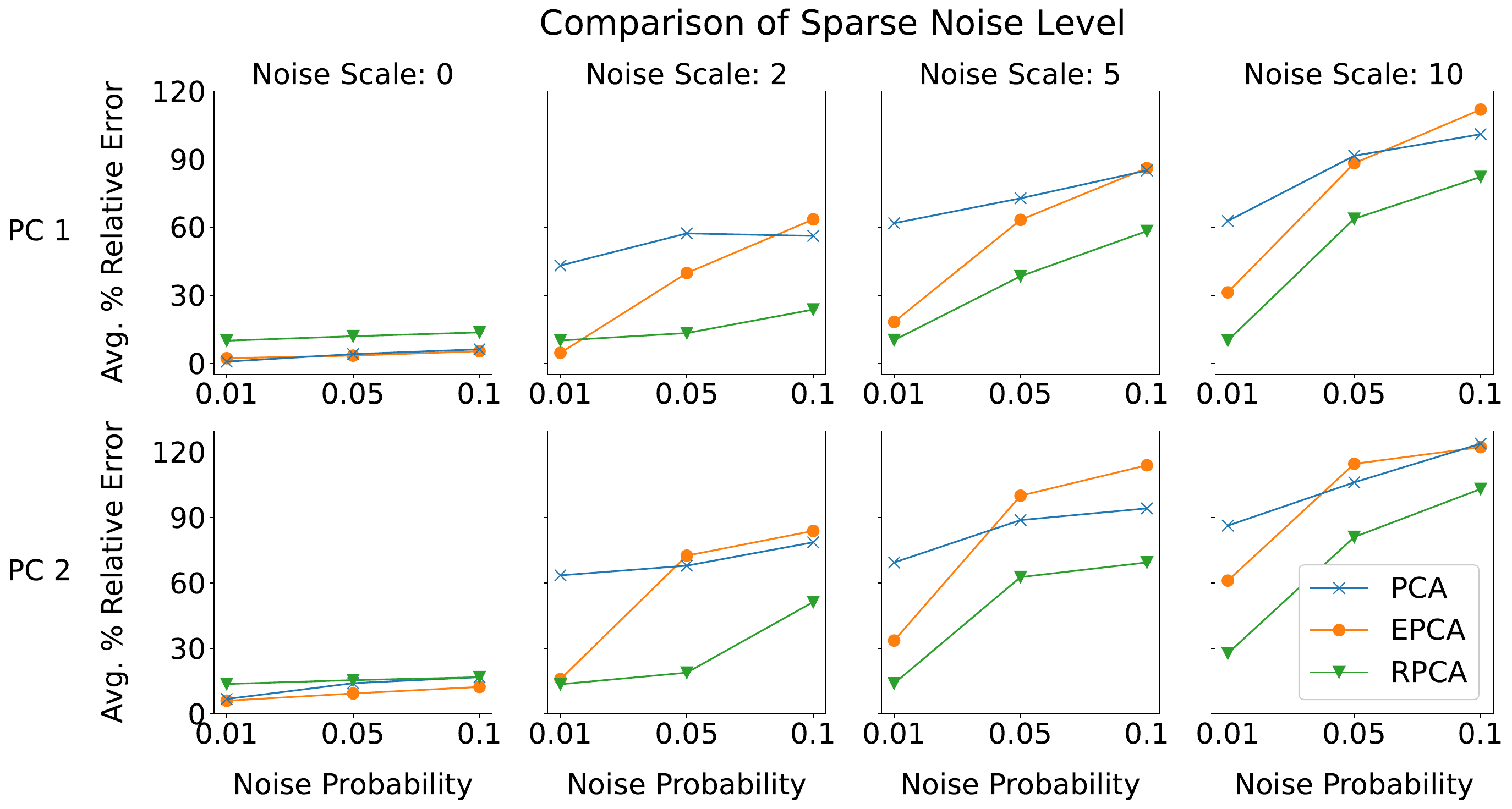}
    \caption{Average percent relative error of predicted principal components from PCA, EPCA, and RPCA for data corrupted with sparse noise of different probabilities and scales.}
    \label{fig:sparse_noise_comparison}
\end{figure}

Figure \ref{fig:sparse_noise_comparison} compares the average performance of PCA, EPCA, and RPCA over datasets corrupted with various levels of sparse noise, determined by the probability $p$ of an entry being corrupt and the scale $c$ of the corruption. As we expect, RPCA performs best on sparse noise, most noticeably as the probability and scale of the noise increases. The exception is when the noise scale is set to $0$, simulating missing data. However, it is possible that the RPCA regularization parameter $\alpha$ is simply not tuned optimally in this case. 
When ${p=0.01}$, EPCA consistently performs second-best. When ${p=0.05}$, EPCA performs second-best in terms of error in the first principal component and slightly worse than PCA on the second principal component. Finally when ${p=.10}$, EPCA generally performs worst of the three methods. Ultimately, RPCA is the preferred choice on sparse data, but EPCA is a competitive choice when our noise probability is very small. 

\begin{figure}
    \centering \includegraphics[width=.5\linewidth]{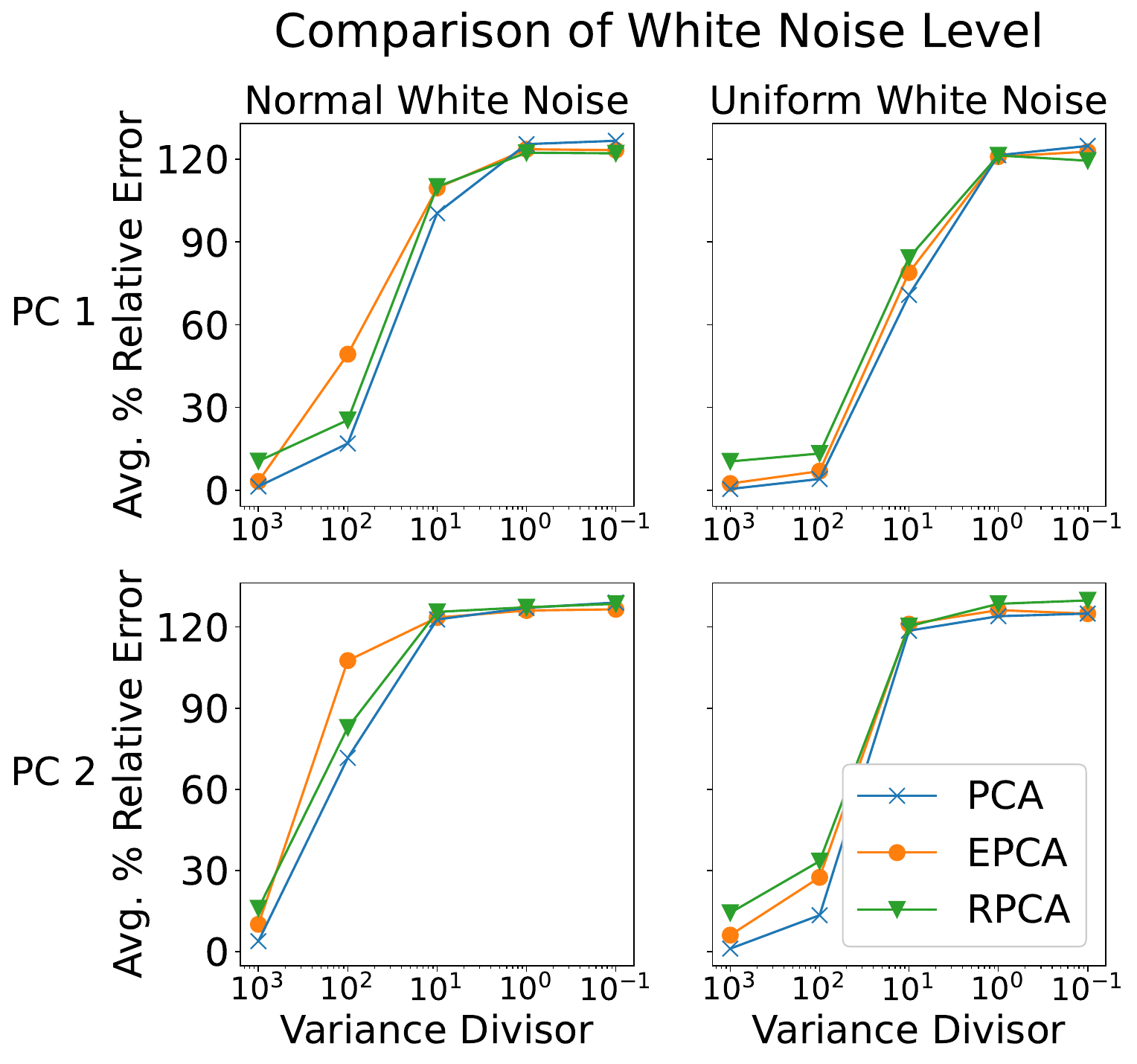}
    \caption{Average percent relative error of predicted principal components from PCA, EPCA, and RPCA for data corrupted with Gaussian and uniform white noise sampled from distributions of various variances.}
    \label{fig:white_noise_comparison}
\end{figure}

Figure \ref{fig:white_noise_comparison} explores the performance of the three methods on data with added Gaussian and uniform white noise sampled from distributions with different variances ${v = \frac{\sigma_1}{f}}$. When the variance divisor $f$ is very large, PCA performs best in terms of the average percent relative error in the first and second principal components. As the variance divisor becomes smaller, PCA, EPCA, and RPCA perform very similarly to one another, but PCA maintains its advantage.

\begin{figure}
    \centering \includegraphics[width=\linewidth]{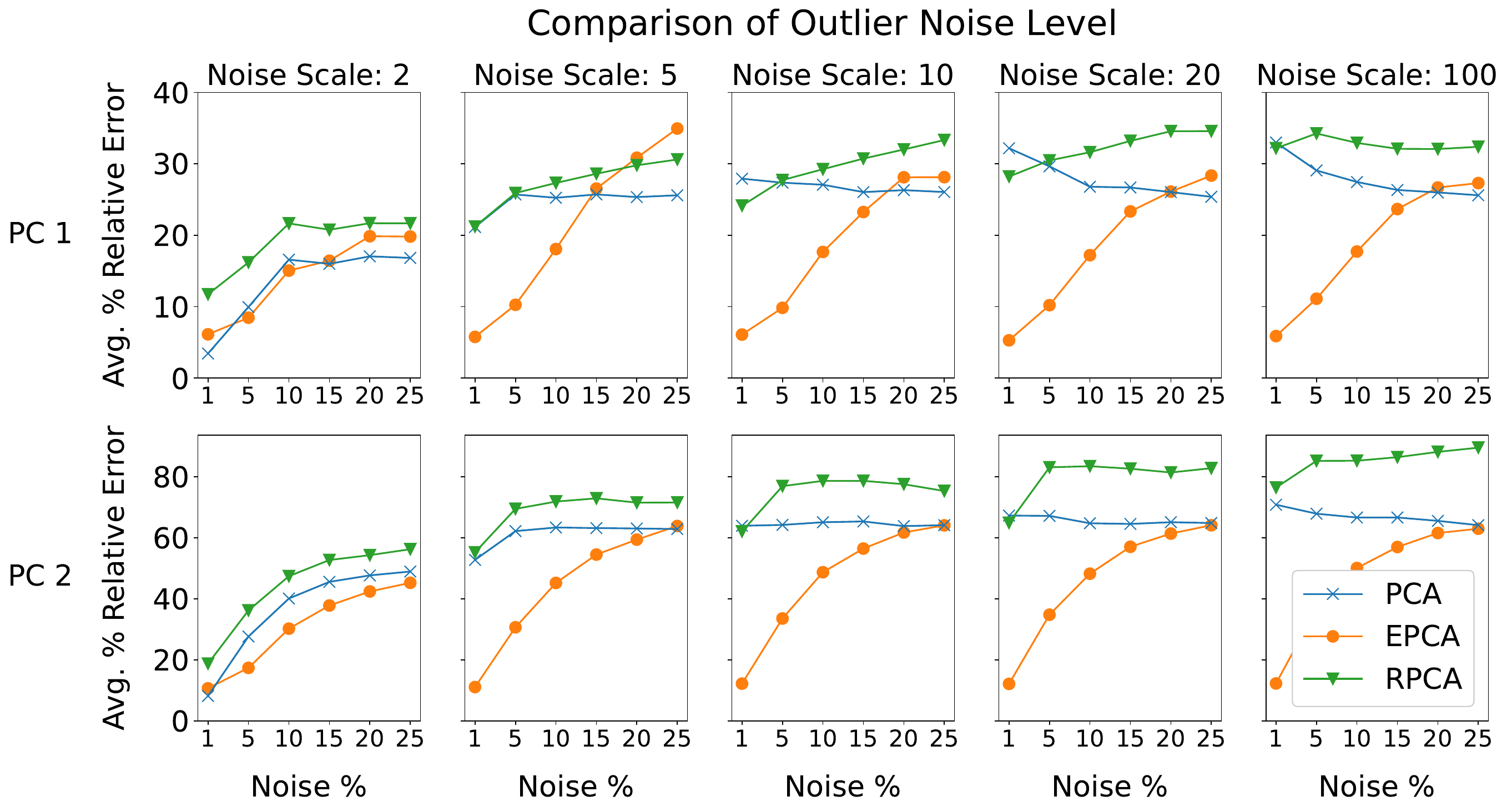}
    \caption{Average percent relative error of predicted principal components from PCA, EPCA, and RPCA for data corrupt with outliers of varied percentages and magnitudes.}
    \label{fig:outlier_comparison}
\end{figure}

Figure \ref{fig:outlier_comparison} displays the performance of PCA, EPCA, and RPCA on data corrupted with outliers at various percentages $s$ and scales $S$. We observe that for both principal components, when the percentage of outliers is $15\%$ or lower, EPCA consistently achieves the lowest average percent relative error, regardless of outlier scale. 
As the percentage of outliers increases, EPCA loses its advantage. 
We conclude that EPCA will generally outperform PCA and RPCA on outlier data, when the percentage of outliers remains small, regardless of the scale of those outliers. 

\subsection{Fixed Level of Noise}
We investigate the performance of PCA, EPCA, and RPCA over datasets with fixed levels of corruption. Recall that all datasets are formed by adding one of four types of noise to seven base datasets 100 times. 
The first aspect of performance we consider is runtime. It is well known that the runtime of RCPA does not scale well with the size of the problem \cite{bouwmans2014robust}. In contrast, since the runtime of EPCA is influenced by the number $B$ and size $n$ of bootstrapped samples, we can mitigate runtime challenges on larger datasets. As PCA is a subroutine of EPCA, we do not expect EPCA to outperform it.

\begin{figure}
     \centering \includegraphics[width=\linewidth]{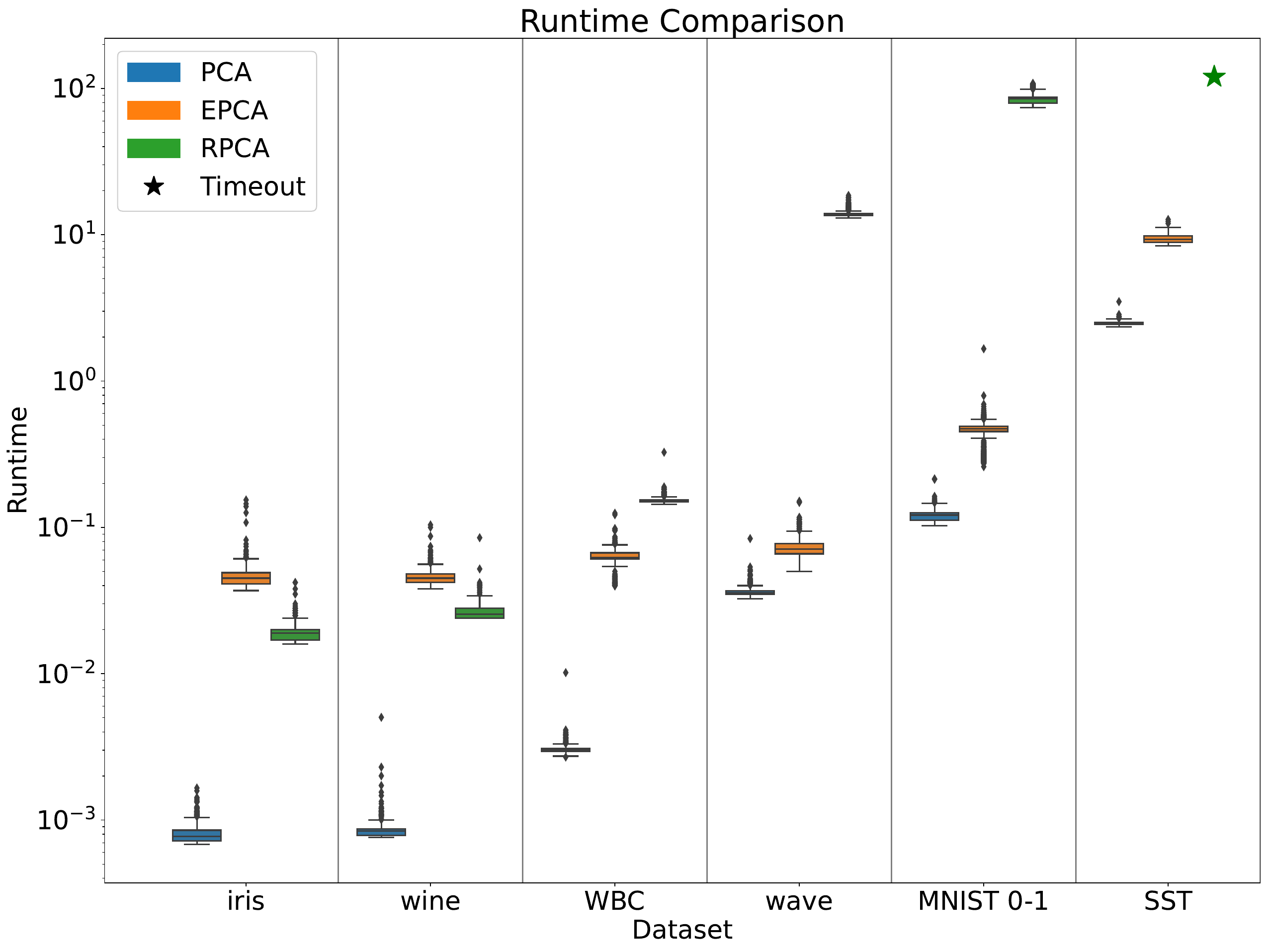}
    \caption{Runtime Comparison of PCA, EPCA, and RPCA. For every corrupted dataset, we create boxplots of the spread of the runtime of $400$ runs of each method. For MNIST, we summarize runtime for the digits $0$ and $1$ together. PCA consistently achieves the fastest runtime. On the smallest datasets, iris, wine, and WBC, RPCA and EPCA run in time on the same order of magnitude, while PCA runs one to two orders of magnitude faster. On the medium-sized datasets, wave and MNIST, RPCA runs two orders of magnitude slower than EPCA, but PCA and EPCA run on the same order of magnitude. Finally, on the largest SST data, RPCA is unable to provide an output, timing out after $120$s, while EPCA and PCA run on the same order of magnitude.}
    \label{fig:runtime_comparison}
\end{figure}

For each of our seven corrupted datasets, we create boxplots of the spread of the runtime of each method, as seen in Figure \ref{fig:runtime_comparison}. We summarize runtime for digits $0$ and $1$ in MNIST together. Across all datasets, PCA achieves the fastest runtime. On the smallest datasets, iris (${\bf X} \in \mathbb{R}^{140 \times 4}$), wine (${\bf X} \in \mathbb{R}^{178 \times 13}$), and WBC (${\bf X} \in \mathbb{R}^{569 \times 30}$), RPCA and EPCA run in time on the same order of magnitude, while PCA runs one to two orders of magniutde faster. On the medium-sized datasets wave (${\bf X} \in \mathbb{R}^{6000 \times 200}$) and MNIST $0$ and $1$ (${\bf X} \in \mathbb{R}^{5923 \times 784}$, ${\bf X} \in \mathbb{R}^{6742 \times 784}$), RPCA runs two orders of magnitude slower than EPCA, but PCA and EPCA run on the same order of magnitude. Finally, on the largest SST data (${\bf X} \in \mathbb{R}^{1726 \times 64800}$), RPCA is unable to provide an output, timing out after $120$s, while EPCA and PCA run on the same order of magnitude. We conclude that on small datasets, RPCA and EPCA have similar runtime. However, unlike RPCA, which is infeasible to run on larger datasets, our method EPCA scales no worse than classical PCA.

\begin{figure}
     \centering \includegraphics[width=\linewidth]{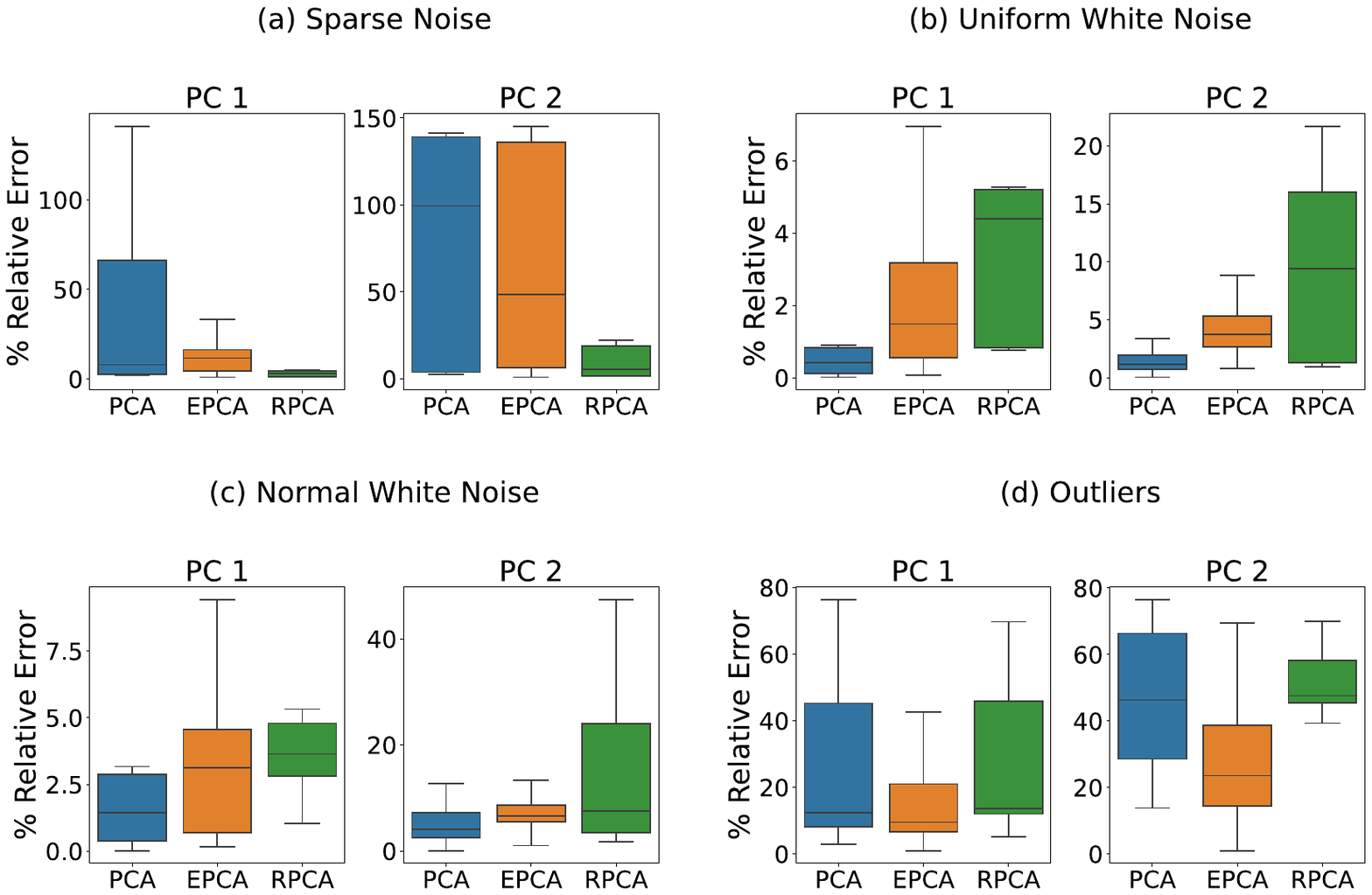}
    \caption{Summary of the performance of PCA, EPCA, and RPCA on noisy data following $100$ trials of random corruption and evaluation on seven datasets. RPCA's boxplots do not include performance metrics from the SST dataset, due to timeout. (a) For sparse noise ($ p = 0.01$, $c=2$), RPCA outperforms the competitors with both lower and tighter IQRs for both principal components. EPCA performs second best with lower median error and tighter IQRs than PCA. (b) and (c) For data corrupted with uniform and Gaussian white noise ($v = \frac{\sigma_1}{1000}$), PCA performs best with tighter and lower IQRs. EPCA achives the second lowest percent relative error on both PCs. (d) For outliers ($s = 5$, $S=5$), EPCA outperforms the competitors with lower median errors for both principal components. EPCA also achieves the tightest IQR on the first and second tightest IQR on the second PC.}
    \label{fig:fullsummary}
\end{figure}

The second aspect of performance that we consider is percent relative error in the predicted first and second components. Figure \ref{fig:fullsummary} shows boxplots of error for each of the three methods for $100$ runs of each type of data corruption over our seven datasets. Outliers in the boxplots have been removed for easier visualization. As expected, on datasets where sparse noise is added, RPCA is able to identify the true principal components with the lowest median percent relative errors and the least variability in its results, as evidenced by tighter interquartile ranges (IQRs). Recall that this performance comes at the cost of a much higher runtime. Though EPCA performs with significantly more error than RPCA, we note that EPCA achieves a similar median error to PCA, as well as a tighter IQR for error in the first PC and both a lower median error and slightly tighter IQR for the second PC.

For uniform white noise, classical PCA outperforms the other methods with both the smallest IQRs and lowest median errors for both principal components. EPCA performs second best in both categories. On data with normal white noise, PCA performs best and EPCA second-best in terms of median error in both PCs. RPCA has the tightest IQR for the first PC, and EPCA for the second. 

Finally, on datasets containing outliers, EPCA outperforms the other methods, achieving a lower median percent relative error for both principal components. EPCA also has the tightest IQR for the first PC and the second-tightest for the second PC. 

\section{Conclusion}

We propose EPCA as a scalable extension of PCA that operates well in the presence of various types of noise and lends itself naturally to uncertainty quantification. Our innovative ensembling of boostrapping and $k$-means clustering allows us to automatically handle the challenges of principal component re-ordering and sign ambiguity in boostrapping PCA. We test the performance of EPCA against RPCA, which is specialized for datasets corrupt with sparse noise, and classical PCA, which should operate well even in the presence of low variance white noise. Further, we carry out a runtime analysis of all three methods on datasets of various sizes.

Overall, EPCA is unable to outperform RPCA on sparsely corrupted data or classical PCA on data with added white noise. 
However, EPCA performs second-best in both noise domains. On datasets containing gross outliers, EPCA significantly outperforms both PCA and RPCA and maintains this performance advantage, regardless of the scale of the outliers, as long as the percentage of outliers remains low. We conclude that although EPCA has a noise domain where it performs best, the method remains useful for all types of data corruption. An added bonus of using EPCA is that unlike classical PCA or RPCA, confidence intervals for the components and the explained variance of those components can be computed naturally. 
Finally, as data size increases, EPCA is significantly faster than RPCA and scales no worse than PCA, making it particularly attractive for the analysis of larger datasets. As each PCA subroutine in EPCA is independent of the rest, the potential for the parallelization of EPCA should be explored as future work.

\vspace{.1 in}
\noindent\textbf{Code Availablity:} https://github.com/OlgaD400/EPCA

\bibliographystyle{plain}
\bibliography{mybib}

\begin{thebibliography}{10}

\bibitem{babamoradi2013bootstrap}
Hamid Babamoradi, Frans van~den Berg, and {\AA}smund Rinnan.
\newblock Bootstrap based confidence limits in principal component analysis—a
  case study.
\newblock {\em Chemometrics and Intelligent Laboratory Systems}, 120:97--105,
  2013.

\bibitem{balasubramanian2002isomap}
Mukund Balasubramanian and Eric~L Schwartz.
\newblock The isomap algorithm and topological stability.
\newblock {\em Science}, 295(5552):7--7, 2002.

\bibitem{benner2015survey}
Peter Benner, Serkan Gugercin, and Karen Willcox.
\newblock A survey of projection-based model reduction methods for parametric
  dynamical systems.
\newblock {\em SIAM review}, 57(4):483--531, 2015.

\bibitem{berkooz1993proper}
Gal Berkooz, Philip Holmes, and John~L Lumley.
\newblock The proper orthogonal decomposition in the analysis of turbulent
  flows.
\newblock {\em Annual review of fluid mechanics}, 25(1):539--575, 1993.

\bibitem{bouwmans2014robust}
Thierry Bouwmans and El~Hadi Zahzah.
\newblock Robust pca via principal component pursuit: A review for a
  comparative evaluation in video surveillance.
\newblock {\em Computer Vision and Image Understanding}, 122:22--34, 2014.

\bibitem{candes2011robust}
Emmanuel~J Cand{\`e}s, Xiaodong Li, Yi~Ma, and John Wright.
\newblock Robust principal component analysis?
\newblock {\em Journal of the ACM (JACM)}, 58(3):1--37, 2011.

\bibitem{cunningham2015linear}
John~P Cunningham and Zoubin Ghahramani.
\newblock Linear dimensionality reduction: Survey, insights, and
  generalizations.
\newblock {\em The Journal of Machine Learning Research}, 16(1):2859--2900,
  2015.

\bibitem{deng2012mnist}
Li~Deng.
\newblock The mnist database of handwritten digit images for machine learning
  research.
\newblock {\em IEEE Signal Processing Magazine}, 29(6):141--142, 2012.

\bibitem{ding2011bayesian}
Xinghao Ding, Lihan He, and Lawrence Carin.
\newblock Bayesian robust principal component analysis.
\newblock {\em IEEE Transactions on Image Processing}, 20(12):3419--3430, 2011.

\bibitem{dony2001karhunen}
R~Dony et~al.
\newblock Karhunen-loeve transform.
\newblock {\em The transform and data compression handbook}, 1(1-34):29, 2001.

\bibitem{drineas2016randnla}
Petros Drineas and Michael~W Mahoney.
\newblock Randnla: randomized numerical linear algebra.
\newblock {\em Communications of the ACM}, 59(6):80--90, 2016.

\bibitem{erichson2016randomized}
N~Benjamin Erichson, Sergey Voronin, Steven~L Brunton, and J~Nathan Kutz.
\newblock Randomized matrix decompositions using r.
\newblock {\em arXiv preprint arXiv:1608.02148}, 2016.

\bibitem{fisher2016fast}
Aaron Fisher, Brian Caffo, Brian Schwartz, and Vadim Zipunnikov.
\newblock Fast, exact bootstrap principal component analysis for p> 1 million.
\newblock {\em Journal of the American Statistical Association},
  111(514):846--860, 2016.

\bibitem{gabrys2006outlier}
Bogdan Gabrys, Bruno Baruque, and Emilio Corchado.
\newblock Outlier resistant pca ensembles.
\newblock In {\em Knowledge-Based Intelligent Information and Engineering
  Systems: 10th International Conference, KES 2006, Bournemouth, UK, October
  9-11, 2006. Proceedings, Part III 10}, pages 432--440. Springer, 2006.

\bibitem{golub1971singular}
Gene~H Golub and Christian Reinsch.
\newblock Singular value decomposition and least squares solutions.
\newblock In {\em Handbook for Automatic Computation: Volume II: Linear
  Algebra}, pages 134--151. Springer, 1971.

\bibitem{goodfellow2016deep}
Ian Goodfellow, Yoshua Bengio, and Aaron Courville.
\newblock {\em Deep learning}.
\newblock MIT press, 2016.

\bibitem{halko2011finding}
Nathan Halko, Per-Gunnar Martinsson, and Joel~A Tropp.
\newblock Finding structure with randomness: Probabilistic algorithms for
  constructing approximate matrix decompositions.
\newblock {\em SIAM review}, 53(2):217--288, 2011.

\bibitem{hannachi2007empirical}
Abdel Hannachi, Ian~T Jolliffe, and David~B Stephenson.
\newblock Empirical orthogonal functions and related techniques in atmospheric
  science: A review.
\newblock {\em International Journal of Climatology: A Journal of the Royal
  Meteorological Society}, 27(9):1119--1152, 2007.

\bibitem{hotelling1938transformation}
Harold Hotelling and Lester~R Frankel.
\newblock The transformation of statistics to simplify their distribution.
\newblock {\em The Annals of Mathematical Statistics}, 9(2):87--96, 1938.

\bibitem{james2013introduction}
Gareth James, Daniela Witten, Trevor Hastie, Robert Tibshirani, et~al.
\newblock {\em An introduction to statistical learning}, volume 112.
\newblock Springer, 2013.

\bibitem{jolliffe2005principal}
Ian Jolliffe.
\newblock Principal component analysis.
\newblock {\em Encyclopedia of statistics in behavioral science}, 2005.

\bibitem{josse2012handling}
Julie Josse and Fran{\c{c}}ois Husson.
\newblock Handling missing values in exploratory multivariate data analysis
  methods.
\newblock {\em Journal de la Soci{\'e}t{\'e} Fran{\c{c}}aise de Statistique},
  153(2):79--99, 2012.

\bibitem{karoui2016bootstrap}
Noureddine~El Karoui and Elizabeth Purdom.
\newblock The bootstrap, covariance matrices and pca in moderate and
  high-dimensions.
\newblock {\em arXiv preprint arXiv:1608.00948}, 2016.

\bibitem{kirby1990application}
Michael Kirby and Lawrence Sirovich.
\newblock Application of the karhunen-loeve procedure for the characterization
  of human faces.
\newblock {\em IEEE Transactions on Pattern analysis and Machine intelligence},
  12(1):103--108, 1990.

\bibitem{kutz2013data}
J~Nathan Kutz.
\newblock {\em Data-driven modeling \& scientific computation: methods for
  complex systems \& big data}.
\newblock OUP Oxford, 2013.

\bibitem{lin2010augmented}
Zhouchen Lin, Minming Chen, and Yi~Ma.
\newblock The augmented lagrange multiplier method for exact recovery of
  corrupted low-rank matrices.
\newblock {\em arXiv preprint arXiv:1009.5055}, 2010.

\bibitem{lorenz1956empirical}
Edward~N Lorenz.
\newblock {\em Empirical orthogonal functions and statistical weather
  prediction}, volume~1.
\newblock Massachusetts Institute of Technology, Department of Meteorology
  Cambridge, 1956.

\bibitem{mika1998kernel}
Sebastian Mika, Bernhard Sch{\"o}lkopf, Alex Smola, Klaus-Robert M{\"u}ller,
  Matthias Scholz, and Gunnar R{\"a}tsch.
\newblock Kernel pca and de-noising in feature spaces.
\newblock {\em Advances in neural information processing systems}, 11, 1998.

\bibitem{milan1995application}
Luis Milan and Joe Whittaker.
\newblock Application of the parametric bootstrap to models that incorporate a
  singular value decomposition.
\newblock {\em Journal of the Royal Statistical Society: Series C (Applied
  Statistics)}, 44(1):31--49, 1995.

\bibitem{sstdata}
NASA~Earth Observations.
\newblock Sea surface temperature.
\newblock Technical report, NASA, 2012.

\bibitem{scikit-learn}
F.~Pedregosa, G.~Varoquaux, A.~Gramfort, V.~Michel, B.~Thirion, O.~Grisel,
  M.~Blondel, P.~Prettenhofer, R.~Weiss, V.~Dubourg, J.~Vanderplas, A.~Passos,
  D.~Cournapeau, M.~Brucher, M.~Perrot, and E.~Duchesnay.
\newblock Scikit-learn: Machine learning in {P}ython.
\newblock {\em Journal of Machine Learning Research}, 12:2825--2830, 2011.

\bibitem{silva2002global}
Vin Silva and Joshua Tenenbaum.
\newblock Global versus local methods in nonlinear dimensionality reduction.
\newblock {\em Advances in neural information processing systems}, 15, 2002.

\bibitem{timmerman2007estimating}
Marieke~E Timmerman, Henk~AL Kiers, and Age~K Smilde.
\newblock Estimating confidence intervals for principal component loadings: a
  comparison between the bootstrap and asymptotic results.
\newblock {\em British Journal of Mathematical and Statistical Psychology},
  60(2):295--314, 2007.

\bibitem{tipping1999probabilistic}
Michael~E Tipping and Christopher~M Bishop.
\newblock Probabilistic principal component analysis.
\newblock {\em Journal of the Royal Statistical Society Series B: Statistical
  Methodology}, 61(3):611--622, 1999.

\bibitem{van2004pca}
S~Van~Aelst and G~Willems.
\newblock Pca based on multivariate mm-estimators with fast and robust
  bootstrap.
\newblock {\em Preprint. MR2085872}, 2004.

\bibitem{van2008visualizing}
Laurens Van~der Maaten and Geoffrey Hinton.
\newblock Visualizing data using t-sne.
\newblock {\em Journal of machine learning research}, 9(11), 2008.

\bibitem{vincent2008extracting}
Pascal Vincent, Hugo Larochelle, Yoshua Bengio, and Pierre-Antoine Manzagol.
\newblock Extracting and composing robust features with denoising autoencoders.
\newblock In {\em Proceedings of the 25th international conference on Machine
  learning}, pages 1096--1103, 2008.

\bibitem{wright2022high}
John Wright and Yi~Ma.
\newblock {\em High-dimensional data analysis with low-dimensional models:
  Principles, computation, and applications}.
\newblock Cambridge University Press, 2022.

\bibitem{zhou2010stable}
Zihan Zhou, Xiaodong Li, John Wright, Emmanuel Candes, and Yi~Ma.
\newblock Stable principal component pursuit.
\newblock In {\em 2010 IEEE international symposium on information theory},
  pages 1518--1522. IEEE, 2010.

\bibitem{zhu2019high}
Ziwei Zhu, Tengyao Wang, and Richard~J Samworth.
\newblock High-dimensional principal component analysis with heterogeneous
  missingness.
\newblock {\em arXiv preprint arXiv:1906.12125}, 2019.

\bibitem{zou2006sparse}
Hui Zou, Trevor Hastie, and Robert Tibshirani.
\newblock Sparse principal component analysis.
\newblock {\em Journal of computational and graphical statistics},
  15(2):265--286, 2006.

\end{thebibliography}
\end{document}